\documentclass[11pt]{article}

\usepackage[preprint]{acl}

\usepackage{times}
\usepackage{latexsym}

\usepackage{microtype}
\usepackage{inconsolata}
\usepackage{booktabs}
\usepackage{graphicx}
\usepackage{mathtools}
\usepackage{amsmath, amssymb}
\usepackage{algorithm}
\usepackage{pgfplots}
\pgfplotsset{compat=1.18}
\usepackage{algorithmic}
\usepackage{enumitem}
\usepackage{hyperref}
\usepackage{tikz}
\usepackage{makecell}
\usepackage[dvipsnames,table,xcdraw]{xcolor}

\usepackage{tcolorbox}

\usepackage[T1]{fontenc}

\usepackage[utf8]{inputenc}

\usepackage{microtype}

\usepackage{inconsolata}

\usepackage{graphicx}

\title{Towards Self-Improving Error Diagnosis in Multi-Agent Systems}

\author{Jiazheng Li$^1$\thanks{Work done at Amazon Alexa AI.} \quad\quad Emine Yilmaz$^{2,3*}$ \quad\quad Bei Chen$^2$ \quad\quad Dieu-Thu Le$^2$\\
  $^1$King's College London \quad\quad $^2$Amazon Alexa AI \quad\quad $^3$University College London\\
  \texttt{jiazheng.li@kcl.ac.uk} \quad\quad \texttt{\{eminey,chenbe,deule\}@amazon.com}}

\begin{document}
\maketitle
\begin{abstract}
Large Language Model (LLM)-based Multi-Agent Systems (MAS) enable complex problem-solving but introduce significant debugging challenges, characterized by long interaction traces, inter-agent dependencies, and delayed error manifestation. Existing diagnostic approaches often rely on expensive expert annotation or ``LLM-as-a-judge'' paradigms, which struggle to pinpoint decisive error steps within extended contexts. In this paper, we introduce \textsc{ErrorProbe}, a self-improving framework for semantic failure attribution that identifies responsible agents and the originating error step. The framework operates via a three-stage pipeline: (1) operationalizing the MAS failure taxonomy to detect local anomalies, (2) performing symptom-driven backward tracing to prune irrelevant context, and (3) employing a specialized multi-agent team (Strategist, Investigator, Arbiter) to validate error hypotheses through tool-grounded execution. Crucially, \textsc{ErrorProbe} maintains a verified episodic memory that updates only when error patterns are confirmed by executable evidence, without the need for annotation. Experiments across the \textsc{TracerTraj} and \textsc{Who\&When} benchmarks demonstrate that \textsc{ErrorProbe} significantly outperforms baselines, particularly in step-level localization, while the verified memory enables robust cross-domain transfer without retraining.
\end{abstract}

\section{Introduction}
Large language model (LLM)-based multi-agent systems (MAS) have demonstrated strong performance across diverse domains, from software engineering \citep{metagpt, chatdev} and web navigation \citep{webshop, yao2022react} to scientific reasoning \citep{aicoscientist}, education \cite{li-etal-2025-two} and tool use \citep{toolformer,gorilla}. Their core appeal is simple: by decomposing a complex task into multiple steps and assigning specialized roles (e.g., architect, engineer, tester) \citep{li-etal-2024-calibrating, autogen, openai2024swarm, yan2025position}, MAS can solve problems that often exceed the capacity of a single LLM call.

\begin{figure}[t]
\centering
\begin{tikzpicture}
\begin{axis}[
    ybar,
    width=\linewidth,
    height=6.0cm,
    bar width=3.5pt,
    enlarge x limits=0.15,
    ymin=0, ymax=80,
    ylabel={Accuracy (\%)},
    yticklabel style={font=\scriptsize},
    ylabel style={font=\scriptsize},
    major grid style={opacity=0.12},
    ymajorgrids=true,
    xtick={1,2,4,5,7,8},
    xticklabels={\scriptsize Agent,\scriptsize Step,\scriptsize Agent,\scriptsize Step,\scriptsize Agent,\scriptsize Step},
    x tick label style={font=\scriptsize, align=center},
    extra x ticks={1.5, 4.5, 7.5},
    extra x tick labels={\scriptsize\textsf{TracerTraj},\scriptsize\textsf{Who\&When Algo},\scriptsize\textsf{Who\&When Hand}},
    extra x tick style={tick style={draw=none}, tick label style={yshift=-1.2em, font=\bfseries\scriptsize}},
    legend style={at={(0.5,-0.25)}, anchor=north, legend columns=3, font=\scriptsize, draw=none, /tikz/every even column/.append style={column sep=0.5em}}
]

\addplot[fill=gray!35, draw=black] coordinates {
    (1, 67.7) (2, 8.7)
    (4, 55.6) (5, 41.3)
    (7, 43.1) (8, 13.8)
};
\addlegendentry{LLM-as-a-Judge}

\addplot[fill=blue!55, draw=black] coordinates {
    (1, 68.5) (2, 34.6)
    (4, 49.2) (5, 58.7)
    (7, 46.6) (8, 27.6)
};
\addlegendentry{\textsc{ErrorProbe}}

\addplot[fill=teal!55, draw=black] coordinates {
    (1, 73.2) (2, 39.4)
    (4, 60.3) (5, 59.5)
    (7, 50.0) (8, 29.3)
};
\addlegendentry{\textsc{ErrorProbe} + Memory}

\end{axis}
\end{tikzpicture}%
\vspace{-1.0em}
\caption{Claude 3.7 Sonnet results on agent attribution and decisive-step localization.
\textsc{ErrorProbe} substantially improves step-level localization over LLM-as-a-Judge while maintaining competitive agent accuracy.
Adding verified memory retains strong performance while enabling reuse across tasks without LLM training.}
\label{fig:intro_oneplot_agent_step}
\vspace{-0.8em}
\end{figure}
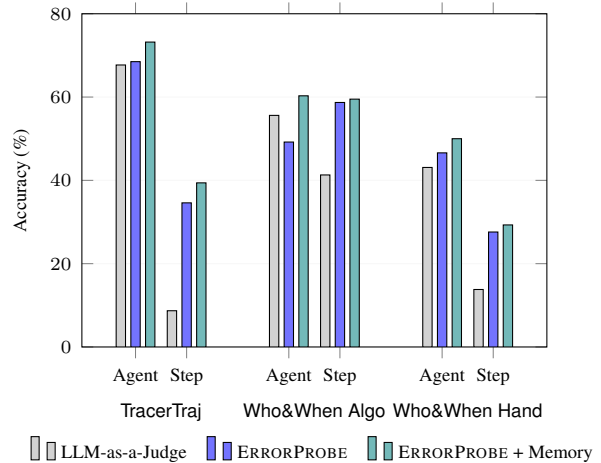

This increased power of MAS creates a practical debugging problem~\citep{zhou-etal-2024-mystery, cemri2025mast, agentracer2025, whoandwhen2025}. Consider a common failure pattern: an architect introduces a subtle specification error, an engineer faithfully implements it, and a tester verifies the wrong behavior. The run may only fail many turns later, after multiple agents have built on the initial mistake. When this happens, practitioners face a central diagnostic question: ``\emph{which agent caused the failure, and at what step did the error originate?}''

In this paper, we study \emph{failure attribution} in MAS--given a multi-agent interaction trace and a failure symptom (e.g., a final wrong answer, an exception, or unmet constraints), the goal is to identify the responsible agent(s) and the originating turn/step (and optionally the failure type). This problem is challenging for several reasons unique to multi-agent settings:
\begin{itemize}[leftmargin=*, itemsep=0.1em, topsep=0pt, parsep=0pt, partopsep=0pt]
  \item \textbf{Long interaction traces:} Production MAS runs can span dozens to over a hundred turns, often exceeding practical context windows \citep{agentracer2025,memgpt,lostinthemiddle}.
  \item \textbf{Delayed manifestation:} Early mistakes may only surface much later, complicating causal attribution \citep{agentracer2025, whoandwhen2025}.
  \item \textbf{Inter-agent dependencies:} Each agent conditions on prior agents' outputs, creating long (and sometimes branching) causal chains \citep{autogen}.
  \item \textbf{Diverse failure modes:} Failures range from specification errors to coordination breakdowns and verification gaps, each requiring different diagnostics \citep{cemri2025mast}.
\end{itemize}

Prior work on failure analysis in LLM-based MAS largely falls into three directions: First, taxonomy-driven studies such as the Multi-Agent System Taxonomy (MAST)~\cite{cemri2025mast} provide a principled lens over diverse failure modes, but constructing and labeling failures requires substantial expert human annotation effort, which is expensive and difficult to scale to the volume and length of real MAS traces \citep{cemri2025mast}. Second, specialized tracers trained on curated supervision can improve attribution performance, but they depend on costly data generation pipelines and may require continual retraining or adaptation as agent frameworks, trace distributions, and error types evolve \citep{agentracer2025}. Third, prompting a general LLM as a judge is attractive for its low overhead and broad applicability, yet current benchmarks show that judge-style methods remain far from reliable for task-specific attribution. Especially for pinpointing the decisive error step in long, delayed-failure traces \citep{whoandwhen2025, ma2025agentfail}. Even training-free transfer mechanisms for judge localization can be sensitive to domain shift \citep{yu2025correct}. These limitations motivate a self-evolving multi-agent diagnosis framework. By emulating human expert debugging: decomposing the diagnosis into specialized roles for backward tracing, hypothesis generation, and tool-grounded verification, such a framework can manage the complexity of long interaction traces and improve attribution accuracy without sacrificing generality across evolving failure modes~\cite{agent_as_judge}.

To tackle the above issues, we present \textbf{\textsc{ErrorProbe}}, a MAS framework for failure attribution. Given the raw interaction history (including inter-agent messages and tool outputs) and a description of the failure symptom (e.g., an exception or incorrect final answer), our framework identifies the responsible agent(s) and the earliest error-inducing step. \textsc{ErrorProbe} performs symptom-driven backward tracing to selectively retrieve causally relevant earlier turns from long interactions, augments the analysis with structured semantic signals from a lightweight detector that operationalizes the MAST failure taxonomy, and maintains a verified memory of reusable error/patch patterns that is updated only when patterns are confirmed through executable evidence (e.g., successful reproduction via code sandboxes) and arbiter verification, improving robustness while avoiding noisy accumulation under distribution shift.

We evaluate \textsc{ErrorProbe} on three benchmarks: \textsc{TraceTraj}~\citep{agentracer2025} and two splits of \textsc{Who\&When}~\citep{whoandwhen2025}, spanning multiple multi-agent frameworks and task settings. Across baselines including LLM-as-a-judge and prior tracing-based approaches, \textsc{ErrorProbe} improves both \emph{agent-level} and \emph{step-level} attribution, with ablations showing that backward tracing, structured failure-mode cues, and verified memory each contribute complementary gains. Finally, we demonstrate that our verified memory design successfully enables cross-domain transfer, allowing patterns learned in one domain to improve attribution in another without the degradation seen in naive approaches. %

Our contributions are as follows:
\begin{itemize}[leftmargin=*, itemsep=0pt, topsep=0pt, parsep=0pt]
    \item We introduce \textbf{\textsc{ErrorProbe}}, a framework for semantic failure attribution in LLM-based multi-agent systems that localizes responsible agent(s) and originating error steps.
    \item We operationalize the \textbf{MAST} taxonomy~\citep{cemri2025mast} into a lightweight detector that scans interaction traces for local anomalies (e.g., misalignment or verification gaps). These structural tags serve as heuristic priors to narrow the search space, providing interpretable, structured failure-mode signals to guide attribution.
    \item We demonstrate effective cross-domain transfer with verified memory, overcoming the brittleness of prior methods that often require retraining or struggle with distribution shift. We show that patterns learned in one dataset (e.g., KodCode) improve diagnosis in another (e.g., TracerTraj), highlighting the importance of verification-aware memory management for robust generalization.
\end{itemize}

\section{Related Work}
\label{sec:related}

\paragraph{Multi-Agent Error Attribution.}
A growing body of work addresses failure attribution in multi-agent systems, identifying responsible agents and error steps within interaction traces. \citet{whoandwhen2025} formalize this problem, highlighting the difficulty of step-level localization. To enable scalable supervision, \citet{agentracer2025} introduce automated data construction via counterfactual replay, while subsequent methods explore spectrum-style analysis \citep{ge2025famas}, hierarchical context decomposition \citep{banerjee2025echo}, and communication efficiency \citep{zhang2024cutcrap}. Complementary research focuses on characterizing failures through taxonomies \citep{cemri2025mast}, root-cause diagnosis benchmarks \citep{ma2025agentfail}, and human-annotated debugging datasets \citep{deshpande2025trail}.
Unlike prior methods that rely on heavy supervision, costly replay, or domain-sensitive caching \citep{yu2025correct}, our work targets semantic diagnosis effective on long, delayed-failure traces by combining backward tracing with structured failure-mode signals.

\paragraph{LLM and Agent Evaluation.}
The ``LLM-as-a-Judge'' paradigm has become a standard for scalable evaluation \citep{zheng2023mtbench}, often enhanced via chain-of-thought or explicit rubrics \citep{liu2023geval,kim2024prometheus}. While recent studies caution against potential biases in automated judges \citep{bavaresco2025judgebench,ye2024justice}, the paradigm has expanded to ``Agent-as-a-Judge,'' which provides process-level feedback for agentic workflows \citep{agent_as_judge}.
However, whereas standard judges typically produce scalar or pairwise rankings, our setting requires causal localization. We therefore move beyond free-form judging to trace-level backward reasoning augmented with semantic verification.

\paragraph{Self-Evolving Agents and Memory.}
Research on self-improving agents leverages verbal reinforcement and memory. Early approaches utilize iterative reflection \citep{madaan2023selfrefine} or episodic buffers \citep{shinn2023reflexion} to refine outputs. To support lifelong learning, systems have adopted skill libraries \citep{wang2023voyager} and memory management architectures \citep{park2023generativeagents,memgpt,shen2025qwenlong}. Recent surveys provide comprehensive taxonomies of agent memory mechanisms across forms, functions, and dynamics \citep{zhang2024memorysurvey,hu2025memorysurvey}. Self-evolving frameworks integrate these components to adapt agent behavior at test-time \citep{liang2024sage,he2025evotest, liu2026selfplay}.
Addressing the challenge of memory corruption in these systems, we propose a verified-before-write update gate. This ensures that only patches supported by verification signals are committed, preventing drift while retaining reusable error patterns.

\begin{figure*}[t]
\centering
\includegraphics[width=0.95\linewidth]{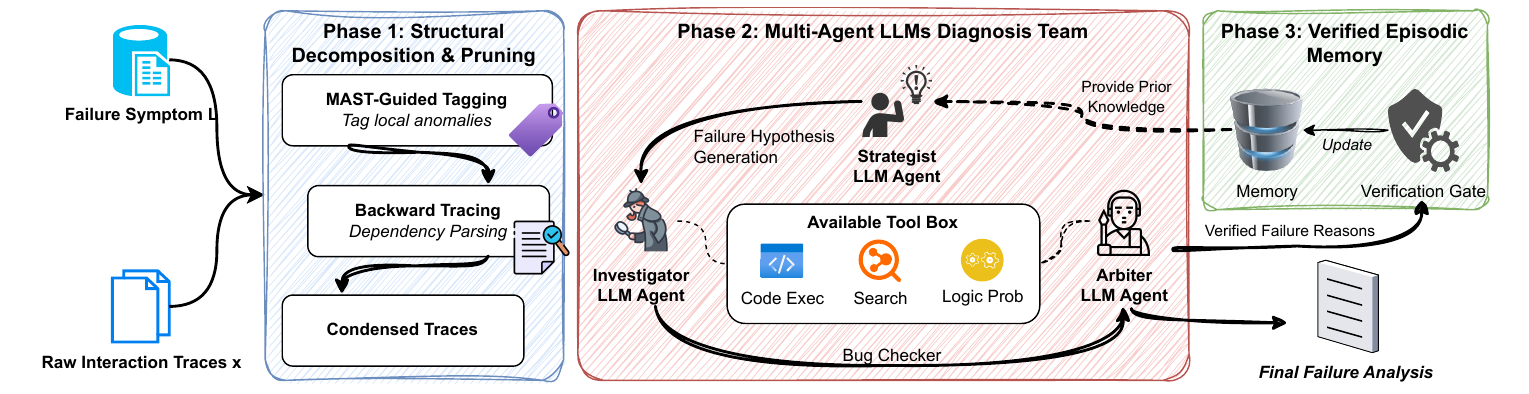}
\caption{Overview of \textsc{ErrorProbe}. The system prunes long traces via dependency parsing, then employs a Strategist-Investigator-Arbiter team to diagnose the root cause, updating memory only upon successful verification.}
\label{fig:arch}
\end{figure*}

\section{Problem Formulation}
\label{sec:problem}

We formalize the problem of \textbf{Multi-Agent Failure Attribution} as a sequential prediction task. Let $\mathcal{T}$ be the space of interaction traces generated by a multi-agent system. These traces consist of a chronological sequence of turn-based interactions, where each step records the sender identity (agent), the recipient, and the semantic content (e.g., natural language messages, code blocks, or tool execution outputs). A task instance is defined as a tuple $(x, y)$, where $x \in \mathcal{T}$ is a failed execution trace containing a sequence of messages and tool outputs $x = [m_1, m_2, \dots, m_L]$, and $y = (a^*, t^*, f^*)$ is the ground-truth attribution tuple consisting of:
\begin{itemize}[leftmargin=1.5em, itemsep=0pt, parsep=0pt]
    \item The \textbf{culprit agent} $a^*$ responsible for the root cause.
    \item The \textbf{decisive error step} $t^* \in [1, L]$, corresponding to the specific interaction $m_{t^*}$ (e.g., a hallucinated instruction, incorrect parameter, or buggy code snippet) where the logic fault was first introduced, as opposed to where the failure symptom became visible.
    \item For the failure mode $f^*$, we specifically adopt the MAST taxonomy~\citep{cemri2025mast}. This framework organizes multi-agent failures into 14 error modes across three hierarchical families:
    \begin{itemize}[leftmargin=1.5em, itemsep=0pt, parsep=0pt]
    \item Specification Issues (FC1): Errors in the initial setup or prompts (e.g., Disobey Task Specification where constraints are violated, or Disobey Role Specification where an agent acts outside its assigned persona).
    \item Misalignment (FC2): Failures in inter-agent coordination (e.g., Ignored Other Agent's Input where context is lost between turns, or Reasoning-Action Mismatch where the output contradicts the agent's internal thought process).
    \item Verification Failures (FC3): Flaws in the quality assurance process (e.g., No/Incomplete Verification where a Tester approves without testing, or Incorrect Verification where valid solutions are rejected).
    \end{itemize}
\end{itemize}

Standard evaluation typically treats this as an i.i.d.\ classification problem, maximizing the likelihood $P(y|x)$. However, real-world debugging is a continuous process where the diagnosis system should learn from past failures. We therefore define a \textbf{Sequential Diagnosis} setting: the system maintains a mutable memory state $\mathcal{M}_i$ at step $i$. Given a stream of failure tasks $\{(x_i, y_i)\}_{i=1}^N$, the model predicts $\hat{y}_i = \mathcal{F}(x_i; \mathcal{M}_{i-1})$ and subsequently updates its state:
\begin{equation}
    \mathcal{M}_i \leftarrow \text{Update}(\mathcal{M}_{i-1}, x_i, \hat{y}_i, \text{Verify}(\hat{y}_i)).
\end{equation}
Our objective is to maximize cumulative attribution accuracy over the stream while minimizing memory size and false positive updates.

\section{The \textsc{ErrorProbe} Framework}
\label{sec:method}

To address the challenges of long contexts and delayed error manifestation, \textsc{ErrorProbe} operates via a three-stage pipeline: (1) Structural Decomposition via the MAST taxonomy, (2) Symptom-Driven Backward Tracing to prune irrelevant context, and (3) a Self-Evolving Multi-Agent Diagnosis loop with verified memory. The overall architecture is illustrated in Figure~\ref{fig:arch}.

\subsection{MAST-Guided Structural Decomposition}
\label{sec:method_mast}
Raw interaction traces are often unstructured and noisy. To provide semantic grounding, we operationalize the \textsc{Mast} taxonomy (defined in Section \ref{sec:problem}) into a lightweight signal detector. Instead of redefining the failure modes here, we focus on detecting their local symptoms.

Rather than asking the model to diagnose $x$ directly, we first parse the trace to extract a structured representation $S_x = \{(\text{agent}_t, \text{role}_t, \text{action}_t)\}_{t=1}^L$. Role ($\text{role}_t$) is the assigned persona (e.g., \textit{Architect}, \textit{Software Engineer}, \textit{Product Manager}) extracted from the system prompt. Action ($\text{action}_t$) is the semantic type of the turn (e.g., \textit{Natural Language Instruction}, \textit{Tool Call}, or \textit{Execution Output}). We then apply a taxonomy-conditioned prompt to tag potential local anomalies. Unlike the global failure symptom provided as input (which indicates that the task failed), these tags identify where specific step-level deviations occurred (e.g., a 'Tool output ignored' warning at Step 5). These weak signals serve as heuristic priors for the subsequent reasoning agents, narrowing the search space from $L$ steps to a smaller set of candidate regions.

\subsection{Symptom-Driven Backward Tracing}
\label{sec:method_tracing}
A core challenge in MAS debugging is the distance between the root cause (e.g., a wrong parameter at $t=5$) and the symptom (e.g., a crash at $t=50$). Processing the full linear history often leads to ``lost-in-the-middle'' phenomena.

Therefore, we implement a \textbf{Backward Tracing} mechanism that reconstructs the causal chain of the failure. Given a failure symptom at step $L$:
\begin{enumerate}[leftmargin=1.5em, itemsep=0pt, topsep=0pt, parsep=0pt]
    \item \textbf{Dependency Parsing:} We construct a dependency graph $G=(V, E)$ where nodes are messages and edges represent information flow (e.g., Agent B cites Agent A's output).
    \item \textbf{Recursive Pruning:} Starting from the symptom node $v_L$, we perform a breadth-first search on incoming edges to identify the \textit{Effective Receptive Field} of the error.
    \item \textbf{Context Compression:} We mask unconnected branches (e.g., parallel sub-tasks that succeeded), retaining only the causal lineage $x' \subset x$.
\end{enumerate}
This condensed context $x'$ focuses the model's attention on the agents and steps that actually influenced the failure, significantly reducing noise for the diagnosis agents.

\subsection{Multi-Agent Diagnosis Architecture}
\label{sec:method_agents}
The diagnosis is executed by a team of three specialized agents designed to mimic a human debugging workflow: hypothesis generation, verification, and adjudication.

\paragraph{The Strategist (Evaluation Lead).} 
This agent acts as the high-level planner. It receives the condensed trace $x'$ and the structural tags from Section~\ref{sec:method_mast}. Its goal is to formulate a set of \textit{Hypotheses} $H = \{(s_j, \text{suspected\_mode})\}$. To do this efficiently, the Strategist queries the long-term memory $\mathcal{M}$ (see Section~\ref{sec:method_memory}) to retrieve $k$ historical failure patterns. In `cold start' scenarios (empty memory) or when no semantically similar patterns are found, the Strategist gracefully degrades to first-principles reasoning, relying solely on the structural MAST tags and trace context. As the system encounters and verifies new failures, this memory naturally populates, progressively shifting the Strategist from zero-shot reasoning to pattern-aware diagnosis. 

\paragraph{The Investigator (Bug Checker).} 
To combat the hallucination of error causes, the Investigator is strictly grounded in tool usage. For every hypothesis $h \in H$, it must produce \textit{evidence} $E_h$. We provide a dedicated toolset:
\begin{itemize}[leftmargin=1em, itemsep=0pt]
    \item \texttt{CodeExec}: Re-runs code snippets in a sandbox to confirm runtime errors.
    \item \texttt{LogicProbe}: Verifies specific pre-/post-conditions (e.g., ``Check if variable $X$ in step $t$ matches the constraints in step $t-5$'').
\end{itemize}
The Investigator cannot simply state ``The code is wrong''; it must generate a diff or an execution log proving the discrepancy.

\paragraph{The Arbiter (Verifier).} 
The Arbiter serves as the final decision gate. It aggregates the hypotheses and their corresponding evidence sets $\{ (h, E_h) \}$, filters out hypotheses where $E_h$ is empty or inconclusive, and outputs the final prediction $\hat{y} = (a, t, f)$ with a confidence score $c \in [0,1]$. The Arbiter also decides whether the diagnosed pattern is novel and robust enough to be committed to memory, acting as the safeguard against memory corruption.

\subsection{Verified Episodic Memory}
\label{sec:method_memory}

To enable self-improvement while mitigating the risk of memory corruption under distribution shift, \textsc{ErrorProbe} maintains a dynamic memory $\mathcal{M}$ of verified failure patterns. Unlike naive caching, we enforce a \textbf{Strict Verification Gate} to ensure only robust diagnoses are retained.

\paragraph{Verification Gate.}
While the Arbiter evaluates evidence to render a verdict for the current task, the Verification Gate acts as a stricter filter for long-term retention. It ensures that only diagnoses with high-confidence, reproducible tool evidence are committed to memory. Formally, a new diagnosis $\hat{y}_t$ and its associated evidence $E_t$ are committed to the memory state $\mathcal{M}_t$ if and only if $\text{Verify}(E_t) \land c_t > \tau$:
\begin{equation}
    \mathcal{M}_t = \mathcal{M}_{t-1} \cup \{(\hat{y}_t, E_t, \sigma_t)\}
\end{equation}
where $\text{Verify}(E_t)$ represents the success of the Investigator's tool-grounded reproduction, $c_t$ is the Arbiter’s confidence, and $\tau$ is a set threshold. This prevents ``hallucinated'' error patterns from polluting the diagnostic history.

\paragraph{Verified Episodic Memory.}
To enable self-improvement while ensuring interpretability, \textsc{ErrorProbe} maintains a dynamic memory $\mathcal{M}$ of verified failure patterns. Each entry $e \in \mathcal{M}$ is stored as a tuple $\langle \mathbf{s}_e, v_e, \sigma_e \rangle$. The structured signature $\mathbf{s}_e$ comprises interpretable features extracted from the trace, including the detected MAST anomaly (e.g., \textit{step\_repetition}), the specific \texttt{tool} or \texttt{api} endpoint involved, the abstracted argument types (\texttt{args}), and context slots (\texttt{ctx}) such as agent role or task domain. Associated with this signature is the diagnosis recipe $v_e$, containing the canonical error pattern and verification guard, and a set of meta-statistics $\sigma_e$ that track recency, frequency, impact, and performance delta.

\noindent \textbf{Relevance Scoring \& Retrieval.}
When diagnosing a new trace $x$, we construct a query signature $\mathbf{s}_x$ and compute a relevance score that combines structural similarity with a quality-weighted RFI-$\Delta$ score:
  \begin{equation}
  \label{eq:retrieval}
  S(x, e) = \underbrace{\text{Sim}(\mathbf{s}_x, \mathbf{s}_e)}_{\text{Structural Match}}
  \times \underbrace{S_{\text{RFI}}(e)}_{\text{Quality Score}}
  \end{equation}
where $\text{Sim}(\cdot)$ computes a weighted match over signature fields (requiring exact match on the MAST tag), and $S_{\text{RFI}}$ is a composite of recency, frequency, impact, and performance delta. To address the cold start problem, we apply a strict similarity threshold $\tau_{\text{ret}}=0.75$. If $\max_{e} \text{Sim}(x, e) < \tau_{\text{ret}}$, retrieval returns an empty set, and the Analyzer defaults to first-principles reasoning using only the MAST taxonomy.

\begin{algorithm}[t]
\small
\caption{Self-Evolving Diagnosis Routine}
\label{alg:diagnosis}
\begin{algorithmic}[1]
\REQUIRE Failed trace $x_t$, Memory $\mathcal{M}_{t-1}$
\ENSURE Diagnosis $\hat{y}$, Updated Memory $\mathcal{M}_t$
\STATE \textbf{Compression:} $x'_t \leftarrow \text{BackwardTrace}(x_t, \text{MastPriors})$
\STATE \textbf{Retrieval:} $\mathcal{P} \leftarrow \text{Retrieve}(\mathcal{M}_{t-1}, x'_t, k)$
\STATE \textbf{Plan:} $H \leftarrow \text{Strategist}(x'_t, \mathcal{P})$
\FOR{hypothesis $h$ in $H$}
    \STATE $E_h \leftarrow \text{Investigator}(h, \text{Tools})$ 
\ENDFOR
\STATE \textbf{Verdict:} $(\hat{y}, c, E) \leftarrow \text{Arbiter}(H, \{E_h\})$
\IF{$c \ge \tau$ \AND $\text{Verify}(E)$}
    \STATE $\mathcal{M}_t \leftarrow \text{Update}(\mathcal{M}_{t-1}, \hat{y}, E)$
\ELSE
    \STATE $\mathcal{M}_t \leftarrow \mathcal{M}_{t-1}$
\ENDIF
\RETURN $\hat{y}$
\end{algorithmic}
\end{algorithm}

\section{Experiments}
\label{sec:exp}

\subsection{Experimental Setup}
\label{sec:exp-setup}

\paragraph{Benchmarks.} 
We evaluate \textsc{ErrorProbe} on three multi-agent failure attribution benchmarks (details in Appendix~\ref{app:benchmarks}). (i) \textbf{TracerTraj}: A code-generation subset of the \textsc{TracerTraj} dataset \citep{agentracer2025}, featuring long-context traces with programmatically injected faults that provide precise, synthetic ground truth. (ii) \textbf{Who\&When-Algo}: An organic failure dataset from \citet{whoandwhen2025} collected from CaptainAgent, a system where agent teams are algorithmically generated and dynamically tailored to each task. (iii) \textbf{Who\&When-Hand}: A complementary set of organic failures from MagenticOne \citep{whoandwhen2025}, representing hand-crafted agent teams. 

\paragraph{Compared Methods.}
We compare four MAS error localization methods, each instantiated with multiple backbone LLMs.
\textbf{LLM-as-a-Judge} prompts a single LLM with the full trace and failure description to directly output $(\hat{a}, \hat{t})$, following prior judge-style work~\citep{zheng2023mtbench,liu2023geval,whoandwhen2025}; we also evaluate additional Who\&When protocols (\texttt{all at once}, \texttt{step by step}, \texttt{binary search}) in Appendix~\ref{sec:appendix_baselines}.
\textbf{Agent-as-a-Judge (baseline)} adopts the tool-augmented framework from \citet{agent_as_judge}, where the evaluator is an autonomous agent equipped with specialized capabilities (e.g., file reading, code localization, and dependency graphing) to inspect the trajectory and verify intermediate states, rather than relying solely on the static trace context.
\textbf{Agent-as-a-Judge (ours)} is our full \textsc{ErrorProbe} pipeline without the Verified Episodic Memory module: we first perform MAST-guided decomposition and backward tracing to compress the trace, then run the Strategist--Investigator--Arbiter team to produce $(\hat{a}, \hat{t})$.
\textbf{Agent-as-a-Judge (ours, with memory)} additionally enables verified episodic memory, where tasks are processed as a stream and new patterns are committed only when verification succeeds. All four methods are compared using Claude 3.7 Sonnet, GPT-OSS-120B, and Qwen 3 32B as backbone models.\footnote{Many traces approach or exceed Qwen 3 32B's context window in the single-judge setting, leading to severe truncation. We therefore omit Qwen3 32B from LLM-as-a-Judge in Table~\ref{tab:main_results} and only report it in Agent-as-a-Judge settings where backward tracing keeps inputs within budget.} 

\paragraph{Tasks and Metrics.}
Following our formulation in Section~\ref{sec:problem}, we evaluate models on two prediction tasks:
(1) \textbf{agent-level attribution}, where the system must identify the culprit agent $a^*$, and
(2) \textbf{step-level attribution}, where it must localize the decisive error step $t^*$.
We report top-1 \emph{Agent} accuracy and \emph{Step} accuracy for each dataset, and their macro-average across the three benchmarks (Table~\ref{tab:main_results}). 
Unless otherwise stated, failure modes $f^*$ are used only as internal supervision signals rather than as an explicit evaluation target.

\begin{table*}[t]
\centering
\small
\resizebox{\textwidth}{!}{%
\begin{tabular}{lcccccccc}
\toprule
& \multicolumn{2}{c}{\textbf{TracerTraj}} 
& \multicolumn{2}{c}{\textbf{Who \& When} (Algo)} 
& \multicolumn{2}{c}{\textbf{Who \& When} (Hand)} 
& \multicolumn{2}{c}{\textbf{Average}} \\
\cmidrule(lr){2-3} \cmidrule(lr){4-5} \cmidrule(lr){6-7} \cmidrule(lr){8-9}
\textbf{Method} & \textbf{Agent} & \textbf{Step} & \textbf{Agent} & \textbf{Step} & \textbf{Agent} & \textbf{Step} & \textbf{Agent} & \textbf{Step} \\
\midrule
\multicolumn{9}{c}{\textit{\textbf{LLM-as-a-Judge}}} \\
\midrule
Claude 3.7 Sonnet & 67.70\% &  8.70\% & 55.60\% & 41.30\% & 43.10\% & 13.80\% & 57.03\% & 21.27\% \\
GPT-OSS-120B      & 70.00\% &  7.90\% & 44.50\% & 18.50\% & 29.30\% & 19.00\% & 47.93\% & 15.13\% \\
\midrule
\multicolumn{9}{c}{\textit{\textbf{Agent-as-a-Judge (baseline)}}} \\
\midrule
Claude 3.7 Sonnet & 59.30\% & 11.10\% & 21.40\% & 47.60\% & 58.60\% & 15.50\% & 46.43\% & 24.73\% \\
GPT-OSS-120B      & 71.10\% & 15.70\% & 18.30\% & 29.40\% & 50.00\% &  8.60\% & 46.47\% & 17.90\% \\
Qwen3 32B         & 65.60\% & 12.60\% & 19.00\% & 39.70\% & 41.70\% & 15.50\% & 42.10\% & 22.60\% \\
\midrule
\multicolumn{9}{c}{\textit{\textbf{\textsc{ErrorProbe} (ours)}}} \\
\midrule
Claude 3.7 Sonnet & 68.50\% & 34.60\% & 49.20\% & 58.70\% & 46.60\% & 27.6\% & 56.33\% & 41.90\% \\
GPT-OSS-120B      & 70.40\% & 20.50\% & 46.80\% & 29.40\% & 56.90\% & 27.60\% & 58.46\% & 25.83\% \\
Qwen3 32B         & 70.90\% & 33.10\% & 38.10\% & 28.60\% & 43.10\% & 25.90\% & 50.70\% & 29.20\% \\
\midrule
\multicolumn{9}{c}{\textit{\textbf{\textsc{ErrorProbe} (ours, with memory)}}} \\
\midrule
Claude 3.7 Sonnet & 73.20\% & 39.40\% & \textbf{60.30\%} & \textbf{59.50\%} & 50.00\% & 29.30\% & 59.60\% & \textbf{42.73\%} \\
GPT-OSS-120B      & 71.70\% & 29.60\% & 49.20\% & 34.10\% & \textbf{59.90\%} & 29.60\% & \textbf{60.27\%} & 31.10\% \\
Qwen3 32B         & \textbf{77.80\%} & \textbf{44.40\%} & 38.90\% & 29.40\% & 46.60\% & \textbf{31.00\%} & 54.43\% & 34.93\% \\
\bottomrule
\end{tabular}%
}
\caption{Main results comparing LLM-as-a-Judge against Agent-as-a-Judge baselines and our methods. \textbf{Bold} indicates best per column. ``Average'' is macro-averaged to weight each benchmark equally.}
\label{tab:main_results}
\end{table*}

\subsection{Main Comparison}
\label{sec:exp-main}

Table~\ref{tab:main_results} summarizes results across the three benchmarks.

\noindent \textbf{Single-pass judging struggles with step localization on long traces.}
Across backbones, LLM-as-a-Judge achieves moderate agent attribution accuracy but substantially lower step-level accuracy, particularly on settings where the root cause may occur far earlier than the final symptom (e.g., TraceTraj Step $<10\%$ for both Claude and GPT-OSS).
This pattern is consistent with the intuition that a single model pass over a long trace tends to overweight late-stage symptoms and underutilize earlier causal evidence.

\noindent \textbf{Tool-augmented agentic evaluation can help, but is protocol-sensitive.}
Agent-as-a-Judge baselines improve step-level attribution in several settings (e.g., Who\&When-Algo Step increases $6.3\%$ for Claude and $8.9\%$ for GPT-OSS), indicating that interactive inspection can surface additional evidence beyond what the raw trace alone provides.
At the same time, agent-level attribution can be brittle under this baseline protocol (e.g., Who\&When-Algo Agent accuracy decreases for both Claude and GPT-OSS), motivating for a more structured diagnostic procedure.

\noindent \textbf{\textsc{ErrorProbe} substantially improves decisive-step localization.}
Introducing MAST-guided decomposition and backward tracing (\textbf{Agent-as-a-Judge (ours)}) yields large gains in step localization across all three benchmarks for Claude and GPT-OSS.
On average, \textsc{ErrorProbe} improves Step accuracy from $20.6\%$ with Claude and from $13.7\%$ with GPT-OSS.
The largest improvements occur on AgenTracer (Claude Step $25.9\%$; GPT-OSS Step $12.6\%$), consistent with \textsc{ErrorProbe}'s goal of recovering earlier causal steps in long trajectories rather than localizing to the final error manifestation.
Agent attribution also improves substantially on Who\&When-Algo (e.g., Claude Agent $27.8\%$; GPT-OSS Agent $28.5\%$), suggesting that evidence-focused tracing helps disambiguate which agent introduced the decisive mistake.

\noindent \textbf{Verified memory provides additional gains, especially for weaker backbones.}
Enabling verified episodic memory further improves performance, with the clearest gains for GPT-OSS-120B.
For Claude, memory yields smaller but consistent improvements in Step accuracy on average, with notable gains on AgenTracer (Step $4.8\%$) and Who\&When-Algo (Step $0.8\%$).
For Qwen3-32B in the agentic settings, memory similarly lifts both Agent and Step accuracy.
Overall, these results support a key design goal of verified-before-write: memory can reuse robust diagnostic patterns while remaining controlled by explicit verification.

The strongest and most consistent benefit of \textsc{ErrorProbe} is improved step-level localization under long, failure-prone trajectories.
Across benchmarks, the results suggest that (i) interactive inspection is helpful but not sufficient on its own, and (ii) structured decomposition, backward causal tracing, and verification-driven memory together yield more reliable attribution than single-pass judging.

\section{Analysis on Memory}
\label{sec:analysis}

\subsection{In-Domain Memory Evaluation}
\label{sec:indomain_memory}

We compare a \textit{Baseline} (memory off) against a \textit{Memory-Augmented} variant across four domains: MBPP and KodCode (coding), GSM8K and MATH (math reasoning).
Because collecting labeled multi-agent failures is costly, we generate trajectories using multiple MAS frameworks and retain only the failed traces to focus analysis on diagnosis under error.
We split traces into train and test partitions, and manually annotate test traces for evaluation.

\begin{table*}[t]
\centering
\small
\renewcommand{\arraystretch}{1.1}
\setlength{\tabcolsep}{4.5pt}
\begin{tabular}{l cc cc cc cc}
\toprule
& \multicolumn{2}{c}{\textbf{MBPP}} & \multicolumn{2}{c}{\textbf{KodCode}} & \multicolumn{2}{c}{\textbf{GSM8K}} & \multicolumn{2}{c}{\textbf{MATH}} \\
\cmidrule(lr){2-3} \cmidrule(lr){4-5} \cmidrule(lr){6-7} \cmidrule(lr){8-9}
\textbf{Setting} & \textbf{Agent} & \textbf{Step} & \textbf{Agent} & \textbf{Step} & \textbf{Agent} & \textbf{Step} & \textbf{Agent} & \textbf{Step} \\
\midrule
\textit{\textbf{\textsc{ErrorProbe} (ours)}} & 87.3\% & 33.3\% & 64.8\% & 47.2\% & 25.0\% & 30.0\% & 68.6\% & 80.0\% \\
\textit{\textbf{\textsc{ErrorProbe} (ours, with memory)}} & \textbf{88.9\%} & \textbf{36.5\%} & \textbf{81.9\%} & \textbf{56.8\%} & \textbf{50.0\%} & \textbf{65.0\%} & 68.6\% & \textbf{85.7\%} \\
\midrule
\textit{Improvement} & \textit{+1.6} & \textit{+3.2} & \textit{+17.1} & \textit{+9.6} & \textit{+25.0} & \textit{+35.0} & \textit{+0.0} & \textit{+5.7} \\
\bottomrule
\end{tabular}
\caption{In-domain Memory Comparison with Curated Error MAS Reasoning Traces.}
\label{tab:indomain_results}
\end{table*}

\paragraph{Efficacy of in-domain memory.}
Table~\ref{tab:indomain_results} shows that enabling verified memory yields consistent improvements across domains.
Averaged over the four domains, memory improves agent attribution by $+10.9$ points and step localization by $+13.4$ points, indicating that diagnosis benefits from reusing previously verified error patterns in addition to per-instance reasoning.

The magnitude of improvement correlates with how repetitive error modes are within a domain. GSM8K exhibits the largest gains (+35.0 Step), consistent with failures that often arise from repeated arithmetic or logic templates. KodCode shows strong Agent gains (+17.1), suggesting recurrent coordination and tool/API misuse patterns. In contrast, MATH sees smaller gains, consistent with more heterogeneous reasoning chains.

\subsection{Memory Scaling}
\label{sec:ablation_memory}

\begin{figure}
    \centering
    \includegraphics[width=\linewidth]{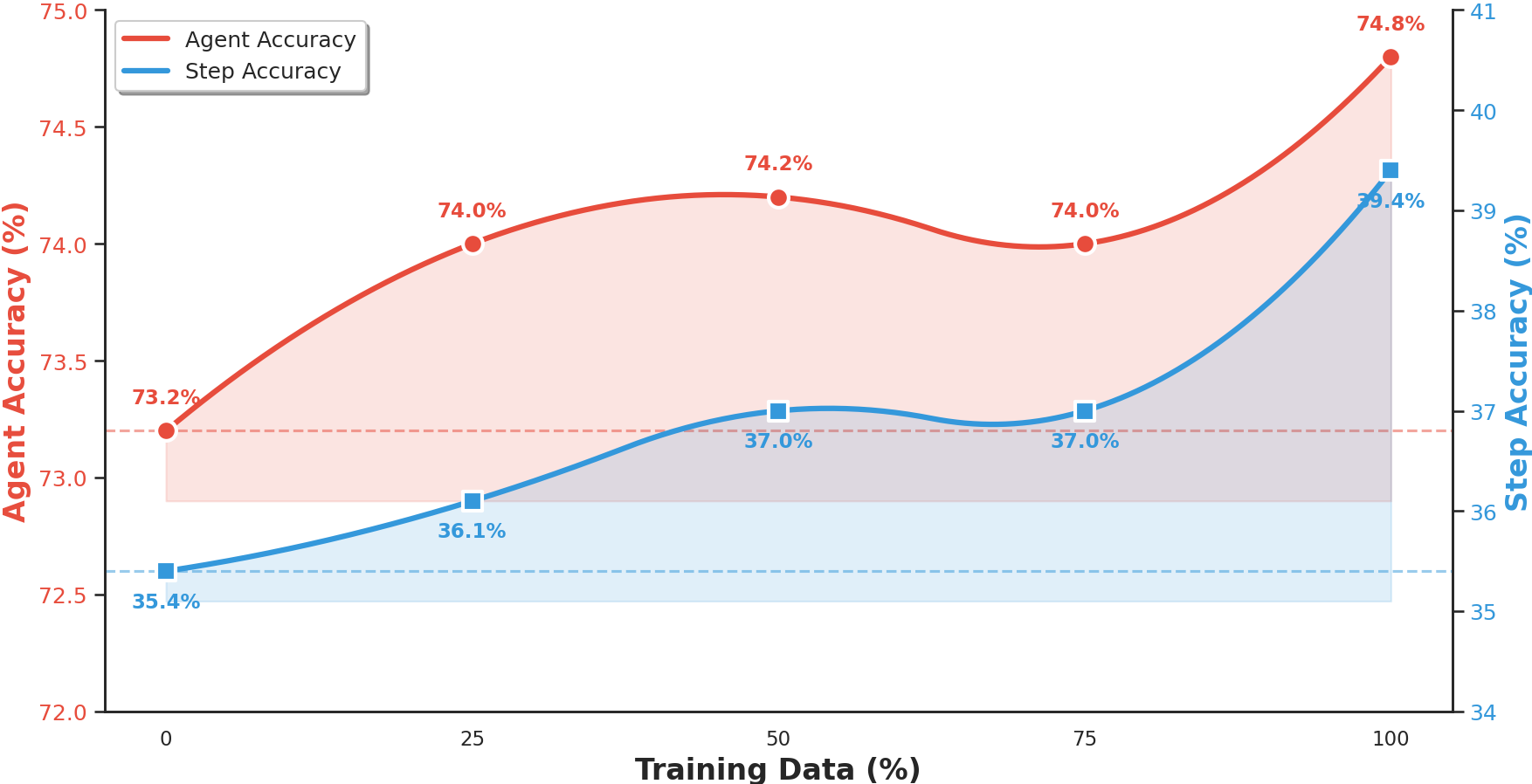}
    \caption{Learning curves for Agent and Step accuracy as memory is populated with increasing portions data.}
    \label{fig:learning_curve}
\end{figure}

Figure~\ref{fig:learning_curve} shows learning curves for Agent and Step Accuracy as memory is populated with increasing training traces from \textit{KodCode}. Both metrics benefit from memory, but with different sensitivities. Agent Accuracy improves $1.6\%$, while Step Accuracy shows more pronounced gains $4.0\%$. This suggests base LLMs have strong priors for agent attribution but struggle with pinpointing the decisive error step; the memory module provides ``error signatures'' that disambiguate the root cause step.

These gains are achieved in an OOD setting, where patterns from \textit{KodCode} are applied to \textit{AgenTracer}. The steady upward trend indicates that the \textit{Verified-Before-Write} gate successfully filters domain-specific noise, retaining only structural motifs that generalize across frameworks.

The steepest Step Accuracy improvement occurs between $75\%$ and $100\%$ data marks the diagnostic ``memory'' is still expanding; further scaling should yield marginal gains for rare failure modes.

\subsection{Case Study: Memory Turns Abstention into Correct Attribution}
\label{subsec:case_study}

In complex traces, baselines often abstain due to ambiguous evidence. Memory augmentation resolves this by retrieving verified diagnostic guards. We illustrate this in the following KodCode trace:

\noindent\fbox{%
    \parbox{\linewidth}{%
        \small
        \textbf{Scenario:} The agent attempts a Shell Sort implementation. \\
        \textbf{Ground Truth:} Agent=\textit{Engineer}, Step=8. \\
        \textit{``The Engineer claims successful Shell Sort implementation, but the file content is truncated.''}
        
        \vspace{0.3em}
        \hrule
        \vspace{0.3em}
        
        \textbf{Baseline Prediction:} \\
        Agent=\texttt{unknown}, Step=$[10]$ \\
        $\rightarrow$ \textcolor{red}{\textbf{Abstention}} (Incorrect)
        
        \vspace{0.2em}
        \textbf{+Memory Prediction:} \\
        Agent=\textit{Engineer}, Step=$[8]$ \\
        $\rightarrow$ \textcolor{ForestGreen}{\textbf{Correct Attribution}} (Gold step $8$)
    }%
}

Here, the baseline abstained due to insufficient explicit evidence. However, the memory-augmented system retrieved a guard advising to ``\textit{verify that all file operations were successfully executed}''. This primed the model to identify the \texttt{reasoning\_action\_mismatch} pattern, correctly attributing the error to unverified claims. For a detailed analysis of the failure modes resolved by memory and the specific mechanisms driving these improvements, see Appendix~\ref{app:extended_analysis}.

\section{Conclusion}
\label{sec:conclusion}

We have introduced \textsc{ErrorProbe}, a framework designed to resolve the challenge of semantic failure attribution in complex multi-agent workflows. By synthesizing symptom-driven backward tracing with a verified episodic memory, our approach effectively filters interaction noise and capitalizes on historical error patterns to enhance diagnostic precision. Empirical evaluations across three diverse benchmarks confirm that \textsc{ErrorProbe} consistently outperforms recent baselines, specifically excelling in the difficult task of temporal error localization. Ultimately, these results underscore that structured, tool-grounded verification is a requisite component for building trustworthy multi-agent systems capable of self-improvement.

\section*{Limitations}

While \textsc{ErrorProbe} offers significant improvements in multi-agent failure attribution, we identify several boundaries to its current scope:

\paragraph{Dependency on Explicit Signals.}
Our symptom-driven backward tracing is predicated on the presence of detectable anomalies (e.g., exceptions, logic inconsistencies, or verifiable constraints). Consequently, ``silent failures'', where agents produce technically valid but semantically incorrect outputs without triggering MAST detectors, remain a challenge. This is a known hurdle in fault localization; future work may integrate test-time oracle feedbacks to expose such latent errors.

\paragraph{Inference Latency.}
The reliability of our multi-agent diagnostic team (Strategist, Investigator, Arbiter) comes at the cost of increased inference compute compared to other non-MAS baselines. While our backward tracing mechanism significantly optimizes context usage, the iterative nature of tool-assisted verification currently constrains applicability in ultra-low-latency production environments. We view this as a necessary trade-off for high-precision debugging in offline or asynchronous workflows.

\paragraph{Backbone Model Diversity.}
Due to the significant computational costs associated with multi-agent evaluation, we limited our experiments to three distinct model families. While we believe this selection provides sufficient evidence of the framework's robustness across different architectures, we were unable to exhaustively validate performance against other high-end proprietary models. Future work should aim to extend this evaluation to a broader spectrum of LLM backbones to further confirm architecture-agnostic generalizability.

\section*{Ethical Considerations}

This work introduces \textsc{ErrorProbe} to assist in the benign debugging of multi-agent systems; we explicitly do not endorse its use for optimizing malicious agent behaviors or circumventing safety measures. While the framework improves diagnostic precision, it relies on explicit failure signals, limiting its utility for optimizing stealthy or ``silent'' malicious deviations. Our experiments utilize established public benchmarks strictly adhering to their respective licenses, and do not involve the collection of new human data.

\section*{Acknowledgments}
We would like to thank Thinh Vinh Ho, Kexin Wang, and Hasan Ferit Eniser from the Alexa AIDo BFA team for their valuable support with the internal experiments.

\bibliography{custom}

\appendix

\section{Experimental Details}
\label{sec:appendix_setup}

\paragraph{Hyperparameters}
\label{sec:appendix_hyperparams}

For each backbone and method, we tune any method-specific hyperparameters (e.g., number of retrieved memory entries $k$, verification threshold $\tau$) on a held-out subset of \textsc{TracerTraj}, and reuse the same configuration across all datasets to avoid test-set overfitting.

All reported numbers are averaged over 5 independent runs with default temperature. Standard deviations across runs were small (typically ${\pm}1.5\%$ for Agent accuracy and ${\pm}2.5\%$ for Step accuracy), so we report median only in Table~\ref{tab:main_results} for clarity. Using a two-proportion z-test, the Step accuracy improvements of \textsc{ErrorProbe} over LLM-as-a-Judge are statistically significant ($p{<}0.01$) across all benchmarks, with effect sizes of +30.7\% (TracerTraj), +18.2\% (Who\&When Algo), and +15.5\% (Who\&When Hand).

Table~\ref{tab:hyperparams} summarizes the key hyperparameters used in \textsc{ErrorProbe}. All values were tuned on a held-out subset of \textsc{TracerTraj} (10\% of traces) and fixed across all evaluation benchmarks.

\begin{table}[h]
\centering
\small
\resizebox{\columnwidth}{!}{%
\begin{tabular}{lcp{5cm}}
\toprule
\textbf{Parameter} & \textbf{Value} & \textbf{Description} \\
\midrule
$k$ & 5 & Number of retrieved memory entries for the Strategist \\
$\tau$ & 0.7 & Confidence threshold for memory commitment \\
$\alpha$ & 0.6 & Weight for semantic similarity in hybrid retrieval (vs.\ frequency) \\
$d$ & 1536 & Embedding dimension for memory keys \\
$T$ & 0.7 & Decoding temperature for LLM calls \\
\bottomrule
\end{tabular}%
}
\caption{Hyperparameter settings for \textsc{ErrorProbe}.}
\label{tab:hyperparams}
\end{table}

\paragraph{Benchmark Details}
\label{app:benchmarks}

We evaluate our framework across three distinct datasets chosen to cover a spectrum of difficulty, from simple logic errors to complex structural misalignment. Table~\ref{tab:benchmark_summary} summarizes their key statistics.

\begin{table*}[h]
\centering
\small
\begin{tabular}{l|ccc}
\toprule
\textbf{Feature} & \textbf{AgenTracer} & \textbf{Who\&When-Algo} & \textbf{Who\&When-Hand} \\
\midrule
\textbf{Source Paper} & \citet{agentracer2025} & \citet{whoandwhen2025} & \citet{whoandwhen2025} \\
\textbf{Size} & 127 & 127 & 58 \\
\textbf{Failure Origin} & Synthetic & Organic & Organic \\
\textbf{Ground Truth} & Programmatic Injection & Algorithm Generated & Manual Annotation \\
\textbf{Primary Domain} & Code Generation & Logic \& Reasoning & Web \& Complex Tasks \\
\bottomrule
\end{tabular}
\caption{Comparison of the multi-agent failure attribution benchmarks used in evaluation.}
\label{tab:benchmark_summary}
\end{table*}

\paragraph{TracerTraj.} This benchmark is derived from the \textsc{TracerTraj} dataset introduced by \citet{agentracer2025}, specifically focusing on the code generation domain (e.g., tasks from MBPP+, KodCode). It evaluates agents within popular frameworks such as \textsc{MetaGPT}, \textsc{OpenHands} (formerly OpenDevin), and \textsc{AutoGen}. Unlike datasets with naturally occurring errors, \textsc{AgenTracer} employs a programmatic fault injection mechanism. A specific ``perturbation operator'' is applied to a ground-truth successful trajectory at step $t$ (e.g., modifying a correct code block to contain a bug or altering a tool argument), thereby generating a failure trace. This methodology ensures the ground truth for failure attribution is absolute by construction, eliminating annotation ambiguity and allowing for precise evaluation of the model's ability to locate the exact ``decisive error'' step.

 \paragraph{Who\&When-Algo.} This subset represents organic failures from the \textsc{Who\&When} benchmark \citep{whoandwhen2025}, focusing on systems with algorithmically generated team structures. The trajectories are collected from \textsc{CaptainAgent} and its adaptive variants (e.g., AG2), where a central ``Captain'' dynamically orchestrates nested teams of specialized agents tailored to specific reasoning and logic queries. The errors in this dataset are non-synthetic, arising naturally during the complex interactions of these dynamically formed teams. Ground truth is established through rigorous human annotation, where experts identify the earliest action in the trace that, if corrected, would prevent the task failure. This dataset tests attribution performance in highly adaptive, fluid multi-agent environments.

\paragraph{Who\&When-Hand.} Serving as a proxy for stable, production-grade systems, this subset from \citet{whoandwhen2025} features organic failures from \textsc{MagneticOne}, a high-quality, orchestrator-based system. Unlike the \textit{Algo} subset, the agentic architecture here is fixed, involving a static set of specialized roles (e.g., WebSurfer, FileSurfer, Coder) coordinated by a central orchestrator. The tasks involve complex, multi-step real-world scenarios such as open-ended web navigation and file manipulation. Ground truth is strictly human-annotated, providing a ``gold standard'' for diagnosing failures in mature, tool-heavy agentic workflows where errors often stem from nuanced interaction breakdowns rather than simple logic faults.

\paragraph{Model Versions}
\label{sec:appendix_versions}
For reproducibility, we report the specific model versions used in our experiments. For Claude 3.7 Sonnet, we used \texttt{claude-3-7-sonnet-20250219}. For GPT-OSS-120B, we used \texttt{gpt-oss-120b-1}. For Qwen3 32B, we used \texttt{qwen3-32b-v1} via the API accessible from the Bedrock. All experiments were conducted between Oct and Dec 2025.

\paragraph{Computational Resources}
\label{sec:appendix_compute}
All experiments were conducted using API-based inference with no local GPU requirements. For a single diagnosis, average inference time for \textsc{ErrorProbe} is approximately 45 seconds for Claude 3.7 Sonnet.

\paragraph{Licensing and Intended Use}
\label{sec:appendix_license}
The evaluation benchmarks used in this work are subject to their original licenses: TracerTraj and AgenTracer are released under Apache 2.0~\citep{agentracer2025}, and Who\&When is released under MIT license~\citep{whoandwhen2025}. We intend \textsc{ErrorProbe} for research on multi-agent system debugging and do not endorse its use for circumventing safety measures or concealing malicious agent behavior.

\section{Further Experiments}
\subsection{Who\&When Protocol Baselines}
\label{sec:appendix_baselines}

We also evaluate three failure attribution protocols from the Who\&When benchmark~\citep{whoandwhen2025}: \texttt{all at once}, \texttt{step by step}, and \texttt{binary search}. Each protocol uses an LLM to predict (i) the responsible agent and (ii) the decisive step that triggered the observed task failure. We instantiate these protocols with three representative backbones (Claude 3.7 Sonnet, GPT-OSS-120B, and Qwen3 32B) and report results on Who\&When (Algorithm-Generated), Who\&When (Hand-Crafted), and TracerTraj-Code. Table~\ref{tab:whowhen_baselines_results} summarizes the results.

\begin{table*}[t]
\centering
\small
\resizebox{\textwidth}{!}{%
\begin{tabular}{lcccccccc}
\toprule
& \multicolumn{2}{c}{\textbf{Who \& When} (Algo)} 
& \multicolumn{2}{c}{\textbf{Who \& When} (Hand)} 
& \multicolumn{2}{c}{\textbf{TracerTraj}} 
& \multicolumn{2}{c}{\textbf{Average}} \\
\cmidrule(lr){2-3} \cmidrule(lr){4-5} \cmidrule(lr){6-7} \cmidrule(lr){8-9}
\textbf{Model} & \textbf{Agent} & \textbf{Step} & \textbf{Agent} & \textbf{Step} & \textbf{Agent} & \textbf{Step} & \textbf{Agent} & \textbf{Step} \\
\midrule
\multicolumn{9}{c}{\textit{\textbf{\texttt{all at once}}}} \\
\midrule
Claude 3.7 Sonnet & 60.32\% & 25.40\% & 39.66\% &  5.17\% & 57.48\% &  3.15\% & 52.49\% & 11.24\% \\
GPT-OSS-120B    & 57.94\% & 26.98\% & 34.48\% &  1.72\% & 55.12\% &  2.36\% & 49.18\% & 10.35\% \\
Qwen3 32B       & 61.11\% & 22.22\% & 41.38\% &  3.45\% & 53.54\% &  3.94\% & 52.01\% &  9.87\% \\
\midrule
\multicolumn{9}{c}{\textit{\textbf{\texttt{step by step}}}} \\
\midrule
Claude 3.7 Sonnet & 23.02\% & 15.08\% & 51.72\% & 17.24\% & 45.67\% &  4.72\% & 40.14\% & 12.35\% \\
GPT-OSS-120B    & 24.60\% & 19.05\% & 44.83\% & 13.79\% & 45.67\% &  3.94\% & 38.37\% & 12.26\% \\
Qwen3 32B       & 23.02\% & 15.87\% & 53.45\% & 20.69\% & 49.61\% &  2.36\% & 42.03\% & 12.97\% \\
\midrule
\multicolumn{9}{c}{\textit{\textbf{\texttt{binary search}}}} \\
\midrule
Claude 3.7 Sonnet & 36.51\% & 11.90\% & 43.10\% & 15.52\% & 53.54\% & 11.81\% & 44.38\% & 13.08\% \\
GPT-OSS-120B    & 34.92\% & 19.84\% & 50.00\% & 13.79\% & 51.97\% &  8.66\% & 45.63\% & 14.10\% \\
Qwen3 32B       & 33.33\% & 17.46\% & 53.45\% & 17.24\% & 55.91\% & 16.54\% & 47.56\% & 17.08\% \\
\bottomrule
\end{tabular}%
}
\caption{Who\&When failure attribution baseline results using three canonical protocols: \texttt{all at once}, \texttt{step by step}, and \texttt{binary search}. We report Agent accuracy and Step accuracy on Who\&When (Algo), Who\&When (Hand-), and TracerTraj.}
\label{tab:whowhen_baselines_results}
\end{table*}

\paragraph{Protocol choice induces a clear agent step trade-off.}
On Who\&When (Algorithm-Generated), \texttt{all\_at\_once} achieves the strongest overall performance across models (57.94--61.11\% Agent; 22.22--26.98\% Step), suggesting that a single-pass judge can succeed when the trace structure is relatively regular and the failure signal is locally salient. In contrast, \texttt{step\_by\_step} substantially lowers agent accuracy on the same dataset (23.02--24.60\% Agent) while retaining moderate step accuracy (15.08--19.05\% Step), indicating that iterative decomposition can reduce reliability in agent attribution even when it remains competitive for step localization.

\paragraph{Hand-Crafted traces benefit from iterative refinement.}
On Who\&When (Hand-Crafted), one-shot judging (\texttt{all\_at\_once}) largely fails to localize decisive steps (1.72--5.17\% Step), despite moderate agent identification (34.48--41.38\% Agent). Both iterative protocols improve step localization substantially, with \texttt{step\_by\_step} reaching up to 20.69\% Step and \texttt{binary\_search} reaching up to 17.24\% Step. This pattern is consistent with Hand-Crafted cases requiring hypothesis refinement across multiple candidate explanations, where the earliest decisive step may be distant from the observed symptom.

\paragraph{TracerTraj-Code is hardest for step localization, and \texttt{binary\_search} is most effective.}
TracerTraj-Code remains challenging for step attribution: \texttt{all\_at\_once} and \texttt{step\_by\_step} yield very low step accuracy across all models (2.36--4.72\% Step). In contrast, \texttt{binary\_search} improves step localization markedly (8.66--16.54\% Step), while maintaining comparable agent accuracy (51.97--55.91\% Agent). This suggests that for longer, code-centric traces, the decisive step is often ``upstream'' of the failure manifestation, and systematic narrowing over the trace is more reliable than either single-shot judgment or purely sequential inspection.

\paragraph{Model choice matters, but less than protocol choice for step localization.}
The best-performing model depends on the dataset and protocol: GPT-OSS-120B performs best on Who\&When (Algorithm-Generated) under \texttt{all\_at\_once} (26.98\% Step), while Qwen3 32B performs best on Who\&When (Hand-Crafted) under \texttt{step\_by\_step} (20.69\% Step) and on TracerTraj-Code under \texttt{binary\_search} (16.54\% Step). However, the variance across protocols is larger than the variance across backbones, especially on Hand-Crafted and TracerTraj-Code where the choice of attribution strategy dominates step-localization performance.

Across all settings, the reproduced Who\&When protocols demonstrate that multi-agent failure attribution is sensitive not only to the judge model but also to the attribution procedure. In particular, \texttt{all\_at\_once} is competitive on more regular traces, while \texttt{binary\_search} is consistently the most effective strategy for step localization on long and code-centric traces.

\subsection{Extended Error Analysis and Mechanisms}
\label{app:extended_analysis}

We analyzed the 32 instances where memory augmentation corrected a previously incorrect baseline prediction. As shown in Table~\ref{tab:memory_error_families}, the distribution reveals that failures related to verification protocols, specifically \texttt{incomplete\_verification} (11) and \texttt{reasoning\_action\_mismatch} (8), are the dominant modes resolved by memory. This suggests that learned guards are particularly effective at enforcing rigorous verification protocols that baseline models frequently overlook.

\begin{table}[h]
\centering
\small
\begin{tabular}{lr}
\toprule
\textbf{Error Family} & \textbf{Count} \\
\midrule
\texttt{incomplete\_verification} & 11 \\
\texttt{reasoning\_action\_mismatch} & 8 \\
\texttt{step\_repetition} & 7 \\
\texttt{context\_loss} & 2 \\
\texttt{fail\_task\_spec} & 2 \\
\texttt{other} & 2 \\
\midrule
\textbf{Total} & \textbf{32} \\
\bottomrule
\end{tabular}
\caption{Distribution of error families in cases where memory enabled correct detection over the baseline.}
\label{tab:memory_error_families}
\end{table}

We attribute this performance gain to three primary mechanisms facilitated by the Error Patch Memory. First, \textbf{Pattern Priming} occurs when retrieved guards prime the Analyzer to identify specific failure signatures (e.g., unverified file operations), effectively directing attention to relevant regions of the context window. Second, \textbf{Span Calibration} leverages positional metadata within memory entries to align predicted error windows to the appropriate trace segments. Finally, \textbf{Taxonomic Guidance} aids in classifying ambiguous failures into actionable categories, significantly reducing model abstention rates by providing a structured decision framework.

\subsection{Illustration of Memory Content}

The following box displays a representative selection of verified failure patterns stored in the episodic memory. These natural language ``guards'' are generated by the Investigator and validated by the Arbiter  to address specific MAST failure modes, such as step\_repetition or reasoning\_action\_mismatch. By retrieving these guards during the diagnosis of new traces, \textsc{ErrorProbe} utilizes historical evidence to prime the model for specific anomalies, thereby resolving ambiguous cases where standard prompts might abstain.

\begin{tcolorbox}[
    colback=gray!10!white,
    colframe=gray!80!black,
    title=\textbf{Illustration of Memory Content},
    boxrule=0.5mm,
    arc=2mm
]
\small
\begin{itemize}
    \setlength\itemsep{0.5em}
    \item Before considering a command execution complete, verify that the entire command was sent and executed successfully. For \texttt{Editor.write} commands, ensure the full file content is included and the command completes with proper JSON structure.
    
    \item After attempting to create a file, verify the file exists by using a tool like \texttt{os.path.exists()} or by listing directory contents. If file creation fails, try an alternative approach or report the error. 
    
    \item Add a tracking mechanism to record what steps have already been completed. Before starting a new thinking or action step, check if it has already been done. For example: if \texttt{'task\_assignment\_complete'} in memory: \texttt{skip\_task\_assignment()}.
    
    \item Before sending a message, verify that the target agent exists in the current role list. If the specified agent is not found, raise a role-specification error or fallback to a valid agent (e.g., use the defined Engineer role instead of ``Alex'').
    
    \item Before issuing a finish command, the Team Leader must verify that the \texttt{solution.py} file actually exists in the workspace and that it passes all tests. Add a guard that checks the file system and runs the test suite, and only proceed to finish when both checks succeed.
    
    \item Add a guard that detects when the Engineer repeats 'thinking' steps without performing any file or code actions for more than two consecutive turns, and forces the next step to be an actionable one (e.g., create a file, write code, or run a test).
    
    \item Add a guard that detects low-complexity (XS) user requests and forces the assistant to respond directly with the solution instead of spawning additional agents.
\end{itemize}
\end{tcolorbox}

\end{document}